# Three-dimensional numerical study of flow characteristic and membrane fouling evolution in an enzymatic membrane reactor

Xu-Qu HU[1,*], Pai-Qing WU[1], Xing-Yi WANG[1], Hai-Cheng ZHANG[1], Jian-Quan LUO[2,†]
1. State Key Laboratory of Advanced Design and Manufacturing for Vehicle Body,
College of Mechanical and Vehicle Engineering, Hunan University,
Changsha 410082, PR China
2. State Key Laboratory of Biochemical Engineering,
Institute of Process Engineering, Chinese Academy of Sciences,
Beijing 100190, PR China

**Abstract:** In order to enhance the understanding of membrane fouling mechanism, the hydrodynamics of granular flow in a stirred enzymatic membrane reactor was numerically investigated in the present study. A three-dimensional Euler-Euler model, coupled with the k-ε mixture turbulence model and the drag function proposed by Syamlal and O'Brien (1989) for interphase momentum exchange, was applied to simulate the two-phase (fluid-solid) turbulent flow. Numerical simulations of single- or two-phase turbulent flow under various stirring speed were implemented. The numerical results coincide very well with some published experimental data. Results for the distributions of velocity, shear stress and turbulent kinetic energy were provided. Our results show that the increase of stirring speed could not only enlarge the circulation loops in the reactor, but it can also increase the shear stress on the membrane surface and accelerate the mixing process of granular materials. The time evolution of volumetric function of granular materials on the membrane surface has qualitatively explained the evolution of membrane fouling.

## 1. INTRODUCTION

Membrane technology, which can be considered as the most energy-saving separation method, has been applied in various industrial processes, *e.g.*, wastewater treatment and biological filtration. It is well known that the membrane fouling is particularly important for constructing a sustainable and high-efficiency operation of membrane technology. To reduce or to avoid the membrane fouling effect, many effective control strategies were proposed over the past decades, such as membrane flushing[1], gas sparging[2], surface modification[3], and high operating shear rate[4]. It is well known that the hydrodynamic properties have great impacts on the effect of membrane fouling. The accurate predictions of flow characteristic and mass transfer occurring are full of importance to design and to evaluate membrane systems[5]. The numerical studies on the membrane separation process in various membrane systems have been the subjects of many researchers during the past decades. A widely investigated problem is the single-phase flow in membrane system, which is fundamental but very helpful for understanding the mechanism of membrane fouling. Torras *et al*. have applied the commercial Computational Fluid Dynamics (CFD) software, Fluent, to investigate the flow characteristic in a rotating disk filtration membrane module under the effects of flow rate, rotation rate and geometry[6]. They showed that the numerically predicated shear stress and pressure variation are in good agreement with analytical solutions. Wang *et al*. used a porous media approach to mimic the membrane structure in a submerged hollow fiber membrane bioreactor[7]. The single-phase hydrodynamic properties were significantly improved by the use of porous media models. Jalilvand *et al.* applied the porous media model in their numerical investigations of high cross-flows in microfiltration process[8]. The hydrodynamic properties (*e.g.*, permeate flux, resistances and shear forces) at different operating conditions were successfully predicated. Recently, Ameur *et al.* have numerically investigated the flow fields and power consumption in a vessel stirred by multistage impeller[9]. They have carefully studied the effects of stirring rate, fluid rheology, vessel size and impeller location on the performance of such stirred system. The studies of single-phase flow, included but not limited to those mentioned above, are full of practical significance for the improvement of membrane performance in various membrane systems. However, the feed solution flowing into membrane systems is usually composed by multiple phases.

---

* Corresponding author, E-mail: huxuqu@gmail.com
† Co-corresponding author, E-mail: jqluo@ipe.ac.cn

During the past decade, an increasing number of researchers have paid their attention on the numerical simulations of multi-phase flow in various reactors. Reza *et al.* constructed a CFD model to solve the momentum and protein concentration continuity equation. The reversible and irreversible membrane fouling resistances were carefully examined in their study. Jong *et al.* have compared the numerical results of Discrete Particle Model (DPM) and Two-Fluid Model (TFM) to the experimental results of a fluidized bed with flat membranes[5]. Their numerical predictions of bubble size distribution during the gas addition coincide well with experimental results. Recently, Trad *et al.* have numerically studied the hydrodynamic characteristics and mixing features of single-phase or two-phase flow in a submerged membrane bioreactor[10]. A compromise operation mode was deduced from their CFD simulations, which was proven to be a balance mode between the preventing of vortex formation and the promoting of homogeneity degree. More numerical studies about membrane performance can be found in Ref.[12, 13]. Despite its paramount importance on the understanding of fouling mechanism, to the best of our knowledge, there are still relatively few studies of membrane fouling evolution with considering of multi-phase flow.

In the present study, we focus on the flow characteristics and membrane fouling mechanism in a widely used enzymatic membrane reactor. We derive a three-dimensional Euler-Euler model, coupled with the k-ε mixture turbulence model and a classical drag function, to describe the two-phase turbulent flow. Firstly, the hydrodynamics of fluid (single-phase) flow in a stirred reactor is numerically studied. The flow characteristics including velocity distribution, shear stress and turbulent kinetic energy are all examined. Numerical simulations of two-phase flow under various stirring speeds are then carried out to study the mixing process of granular flow. The distribution evolution of granular materials on the membrane surface has also been carefully studied to enhance the understanding of membrane fouling mechanism.

## 2. METHODS AND MODELS
## 2.1 PROBLEM DESCRIPTION

Our study is based on a magnetically stirred dead-end cell (Amicon 8050, Millipore, USA), which has been widely used in experiments of membrane filtration and enzymatic reaction[13]. As shown in Fig. 1 (left), a flat membrane with effective area of about 15.9 $cm^2$ (4.5 *mm* diameter) is fixed on the substructure of cylindrical housing. An inverted T-shaped blade is placed near the membrane, which can be rotated to generate shear rate on the membrane surface. Filling the cell with nitrogen gas can provide a nearly constant pressure. The feed solution is pressure-driven to flow towards the membrane surface. Because of the effect of membrane filtration, the granular pollutants are separated from the feed solution and stay in the container.

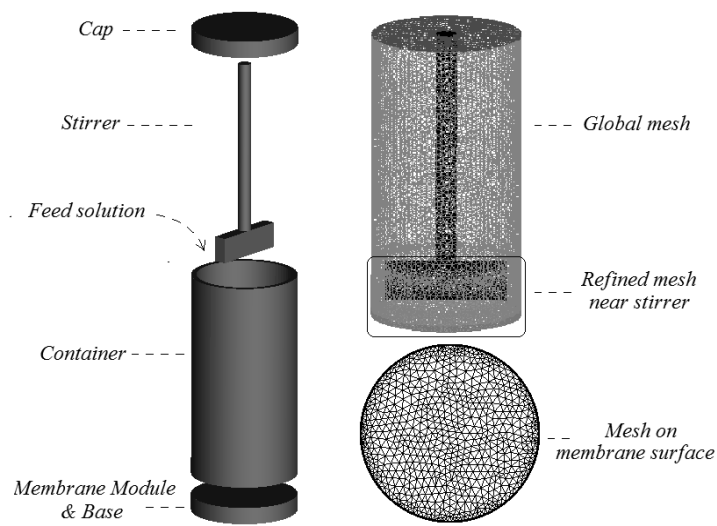

Fig. 1. Schematic of the enzymatic membrane reactor (left) and the corresponding mesh (right).

We assume that the feed solution is composed by multiple phases, in which the main phase of water accounts for 99% volumetric fraction, while the other 1% (secondary phase) is granular pollution material of density 2000 $kg/m^3$. The average diameter of granular material ($D_p$) is 0.05 $mm$. The ultrafiltration membrane settled at the end of reactor is RC10-M or RC05-M (PLGC, Millipore, USA) in our study. The constant rotation speed of the T-shaped blade is assumed as 60, 150 and 250 rounds per minute (*rpm*), respectively.

## 2.2 EULER-EULER APPROACH

The feed solution consisting of a little amount of granular materials (1% in volumetric fraction) dispersed in pure water is considered as interpenetrating continua. The Euler-Euler approach has been widely applied for simulating of unsteady multi-phase flow. The Euler-Euler multiphase in present study takes the solid-solid momentum exchange into account, which cannot be neglected in high-concentration mix flow. We consider that the two phases occupy different volumes of space and apply the Euler-Euler approach to describe this two-phase (fluid-solid) solution. The subscript *i* is used to distinguish the two phases from each other, while *i=s* is for solid and *i=l* is for liquid. The volumetric fractions of both phases ($\alpha_i$) are continuous functions of space and time, and their sum should equal to one,

$$\sum \alpha_i = \alpha_l + \alpha_s = 1. \tag{1}$$

As the solid materials take up only a very small fraction in the mixed phase, we define the liquid as carrier phase. Both phases are modeled as continuum. The governing equations for mass conservation is as,

$$\frac{\partial}{\partial t}(\alpha_i \rho_i) + \nabla \cdot (\alpha_i \rho_i \mathbf{u}_i) = 0, \tag{2}$$

where $\rho$ and **u** are the density and the velocity, respectively. The governing equations for momentum conservations can be written as,

$$\frac{\partial}{\partial t}(\alpha_l \rho_l \mathbf{u}_l) + \nabla \cdot (\alpha_l \rho_l \mathbf{u}_l \mathbf{u}_l)$$
$$= -\alpha_l \nabla p + \nabla \cdot (\alpha_l \bar{\bar{\tau}}_l) - K_{ls}(\mathbf{u}_l - \mathbf{u}_s) + \mathbf{F} \tag{3}$$

$$\frac{\partial}{\partial t}(\alpha_s \rho_s \mathbf{u}_s) + \nabla \cdot (\alpha_s \rho_s \mathbf{u}_s \mathbf{u}_s)$$
$$= -\alpha_s \nabla p + \nabla p_s + \nabla \cdot (\alpha_s \bar{\bar{\tau}}_s) - K_{ls}(\mathbf{u}_l - \mathbf{u}_s) + \mathbf{F} \tag{4}$$

where *p* is the pressure, ***F*** is the external forces acting on the systems, and $K_{ls}$ represents the interphase momentum exchange term. According to Newton's law of viscosity, we can get the viscous stress tensor for both phases,

$$\bar{\bar{\tau}}_i = \left(-\frac{2}{3}\mu_m\right)(\nabla \cdot \mathbf{u}_i)\bar{\bar{I}} + 2\mu_i \bar{\bar{S}}_i, \tag{5}$$

where $\mu$ is the dynamic viscosity, $\bar{\bar{I}}$ is a unit tensor, and $\bar{\bar{S}}$ is the stain rate tension. For the phase of Newtonian fluid, the expansion viscosity $\lambda$ that measures the difference between the thermodynamic and mechanical pressures is about three times larger than $\mu$, while it is zero for the solid phase according to the Stokes' assumption. Although there are plenty of models for dynamic viscosity for solid phase, we have used the widely accepted one proposed by Syamlal *et al.*[15, 16]. It can be expressed as,

$$\mu_{s,kin} = \frac{4}{5}\alpha_s^2 \rho_s D_p g_0 (1+e)\sqrt{\frac{T}{\pi}}$$
$$+ \frac{\alpha_s \rho_s D_p \sqrt{T\pi}}{6(3-e)}\left(1 + \frac{2}{5}(1+e)(3e-1)\alpha_s g_0\right), \tag{6}$$

where $g_0$ is the function of radical distribution, *T* is the thermodynamic temperature of solid material, and *e* represents the restitution coefficient. More details about the theory for dynamic viscosity of solid phase are available in Ref. [11,15, 16].

## 2.3 INTERPHASE MOMENTUM EXCHANGE

For this granular flow, the fluid-solid interaction is mainly achieved through momentum transfer between the two phases. The interphase momentum exchange term ($K_{ls}$), which can be calculated by specifying drag functions, is a key factor in the simulating of granular flows. In present study, we have applied the drag function proposed by Syamlal and O'Brien[15], in which the correlation for momentum exchange coefficients has the following form,

$$K_{ls} = \frac{3}{4}\frac{\alpha_s \alpha_l \rho_l}{v_{r,s}^2 D_p} C_D \left(\frac{\text{Re}_s}{v_{r,s}}\right) |\mathbf{u}_s - \mathbf{u}_l|, \tag{7}$$

where the relative *Reynolds* number *Re*$_s$ is defined as,

$$\text{Re}_s = \frac{\rho_l D_p |\mathbf{u}_s - \mathbf{u}_l|}{\mu_l}. \tag{8}$$

The drag function for a single granular particle $C_D$ has the following form,

$$C_D = \left(0.63 + \frac{4.8}{\sqrt{\text{Re}_s / v_{r,s}}}\right)^2. \tag{9}$$

The terminal velocity correlation for the solid phase $v_{r,s}$ is as,

$$v_{r,s} = \frac{1}{2}\sqrt{(0.06\,\mathrm{Re}_s)^2 + 0.12(2B-A) + A^2} \\ + \frac{A}{2} - 0.03\,\mathrm{Re}_s \quad , \tag{10}$$

where $A = \alpha_l^{4.14}$ and $B = 0.8\alpha_l^{1.28}$.

## 2.4 K-ε MIXTURE TURBULENCE MODEL

As the mixed phase is stirred by high-speed blade rotations, the *Reynolds* number for the flow in reactor is relatively high. The effect of turbulent flow needs to be taken into account in modeling[17]. The standard *k-ε* turbulence model, which is initially proposed by Launder and Spalding (1972), is based on two transport equations for the turbulence kinetic energy ($k$) and its dissipation rate ($\varepsilon$). Its validity for numerical simulations of turbulence flows has been widely proven during the past several decades, *e.g.*, in Ref. [6, 18]. We have applied the *k-ε* mixture turbulence model in present study, which can be expressed as follows,

$$\frac{\partial}{\partial t}(\rho_m k) + \nabla \cdot (\rho_m \mathbf{u}_m k) = \nabla \cdot \left(\frac{\mu_{t,m}}{\sigma_k}\nabla k\right) + G_{k,m} - \rho_m \varepsilon, \tag{11}$$

$$\frac{\partial}{\partial t}(\rho_m \varepsilon) + \nabla \cdot (\rho_m \mathbf{u}_m \varepsilon) \\ = \nabla \cdot \left(\frac{\mu_{t,m}}{\sigma_k}\nabla \varepsilon\right) + \frac{\varepsilon}{k}(C_{1\varepsilon}G_{k,m} - C_{2\varepsilon}\rho_m \varepsilon) \quad , \tag{12}$$

where $\rho_m$, $\mathbf{u}_m$, $\mu_{t,m}$, $G_{k,m}$ represent the density, velocity, turbulent viscosity and production of turbulence kinetic energy of the mixed phase, respectively. They are defined as follows,

$$\rho_m = \sum_{i=1}^{2}\alpha_i \rho_i \;,\; u_m = \frac{\sum_{i=1}^{2}\alpha_i \rho_i \mathbf{u}_i}{\rho_m} \;,\; \mu_{t,m} = \rho_m C_\mu \frac{k^2}{\varepsilon} \;, \\ G_{k,m} = 2\mu_{t,m}\overline{\overline{S}}:\nabla \mathbf{u}_m. \tag{13}$$

The other coefficients are defined as $C_{1\varepsilon}$=1.44, $C_{2\varepsilon}$=1.92, $C_\mu$=0.09 and $\sigma_k$=1.3. They were all determined from experimental results, and the accuracy has been validated for a wide range of turbulent flows[6, 18].

## 2.5 NUMERICAL IMPLEMENTATION

We apply the Multiple Reference Frame (MRF) approach to model the stirring blades. Under the moving frame, the flow around the stirring blade can be simplified as a steady-state problem[18]. For the ultrafiltration membranes, we use the porous medium model to simulate its physical behavior that allows water to pass through it but presents as a barrier for granular particles. We set the porosity of porous medium to 0.0001 so that the granular materials with diameters of 0.05 *mm* cannot penetrate through the membrane surface. The effect of membrane resistance is described by the Darcy's law, for which the coefficients are derived from our experimental results. In the process of operations, we consider that a constant pressure (200 *kPa*) is applied at the inlet, and the outlet is operated with atmospheric pressure.

As shown in Fig. 1 (right), the flow domain is discretized into a finite number of contiguous control volumes by an unstructured mesh of about 272000 tetrahedral elements. These governing equations, which are discretized by the second-order upwind scheme, can be obtained from the finite volume method by Fluent code. We have checked the influence of mesh size using a mesh twice as small as that described above and have not found significant influence on the results of flux. In order to ensure the calculation precision, we have used a refined mesh at the region surrounding the blades. A time step of $0.001s$ is applied for the time integration.

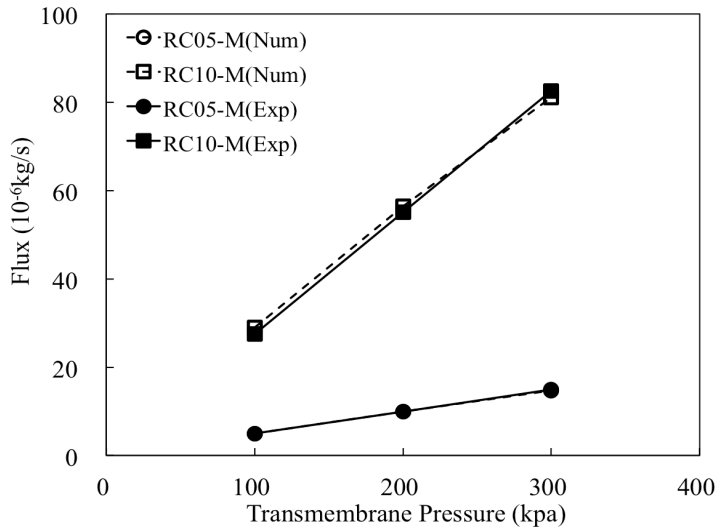

Fig. 2. Comparison of numerical and experimental results of flux to transmembrane pressure.

## 2.6 VALIDATION

The flux variation with respect to trans-membrane pressure (TMP) for many types of ultrafiltration membranes has been experimentally investigated in our former study[13]. We found that, when absorption fouling or pore blocking stays weakly, the total filtration resistance remains constant and the flux increases linearly with TMP. This linear function of flux to TMP can be a good reference for the validations of numerical models. We have studied the pressure-driven water flow in a non-stirred reactor to validate our numerical model. The flux predicated by the present numerical model is compared with published experimental data[14] in Fig. 2. We can find that the numerical and experimental results coincide well with each other for both RC10-M and RC05-M membranes. This implies our numerical models, as well as the porous medium approach that we use, are suitable and accurate for modeling of this membrane system.

## 3. RESULTS AND DISCUSSIONS

We apply the numerical model to simulate the stirred complex flow in an enzymatic membrane reactor. The stirring speed is considered as 250, 150 and 60 *rpm*, respectively. We firstly study the single-phase flow of pure water to examine the flow characteristic in a stirred reactor. The membrane fouling evolution is then studied by numerical simulating of granular flows, in which the fluid-solid phases consist of 99% water and 1% granular solid particle as described above. The following results are all at the steady states.

## 3.1 FLOW CHARACTERISTICS

As shown in Fig. 3 (b), the flow velocity at the membrane center is much lower than that in the marginal area. The rotational speed of fluid can be obviously accelerated by the increasing of stirring speed. In principle, a stronger stirring speed can induce higher shear rate, which makes the surface adhesion of pollution material on membrane harder to form. This result confirms the experimental fact that a higher stirring speed is often beneficial to control or to reduce the effect of membrane fouling.

Flow characteristic can strongly affect the separation efficiency of membrane systems. In this section, we numerically study the single-phase flow in a stirred reactor. Fig. 3 shows the velocity distributions on vertical and horizontal (i.e. membrane surface) planes. We can see that there are two obvious circulation loops existing at both sides of driven shaft, which is in good agreement with previous experimental and numerical studies[11,17,19]. The main reason for these circulation loops is the vortex formation. Due to the blade rotations, the fluid around the blades is stirred, and therefore, centrifuged towards the sidewall of container. This radial jet blocked by wall boundary transforms to two main tributary flows: the upstream flow and the downstream one. Since the fluid is continuously pumped into container, the upstream flow will meet the inlet flow in the middle of container, which can merge to form vortex. The vortex circulating near the shaft takes on circulation loops. It is of interest to notice in Fig. 3 that, the area of circulation loops can be obviously enlarged by the increase of stirring speed. This implies that a higher stirring speed makes less flow concentration near the membrane, which is beneficial for preventing membrane fouling. For the downstream flow, it moves towards the base of container. When it reaches the membrane surface, it partially transforms to the permeating flow since the membrane is modeled as porous media. However, most of it stays inside the container and converges to the rotating flow on the membrane surface. Fig. 4 shows the variation of shear stress along a radial line on the membrane surface. The shear stress is relatively low at the central area, but it increases rapidly with the increasing (or decreasing) of radial position. This radial non-uniformity of shear stress is mainly caused by the inhomogeneous distribution of velocity. As shown in Fig. 3 (b), the velocity distribution at the membrane center implies a low velocity gradient, which is just the reason for the low shear stress. It shows the velocity gradient is negligible for a low stirring speed but notable for high stirring speeds. Accordingly, we can also find in Fig. 4 that the increasing of stirring speed can significantly enlarge the shear stress on the membrane surface.

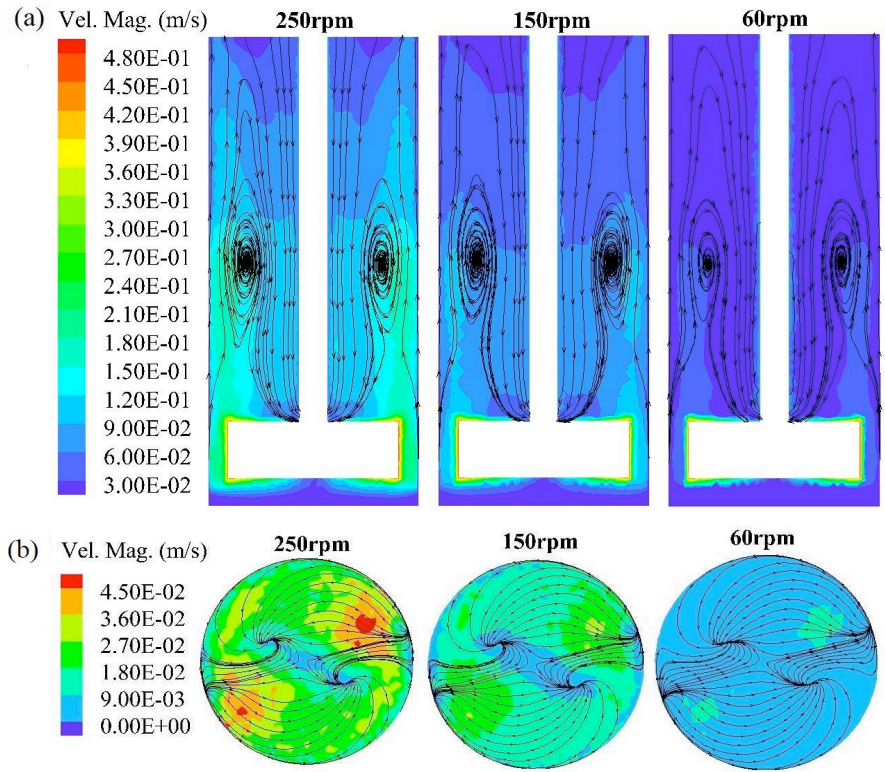

Fig. 3. Velocity distribution on verical (a) and horizontal (b) planes in a stirred reator.

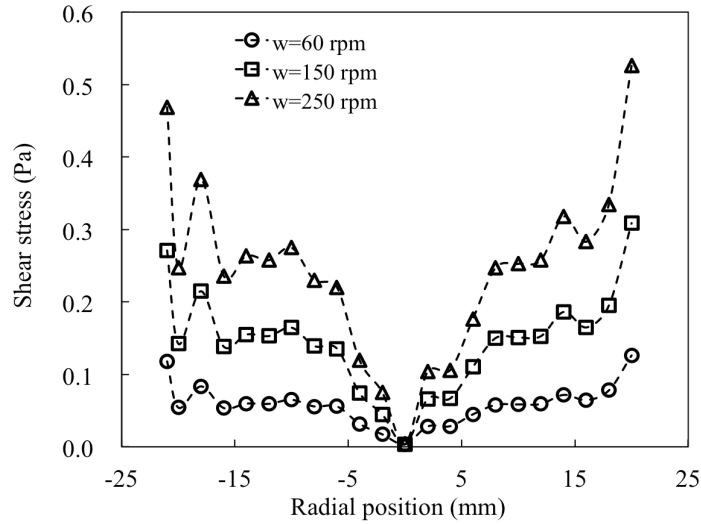

Fig. 4. Variation of shear stress along a radial line on membrane surface.

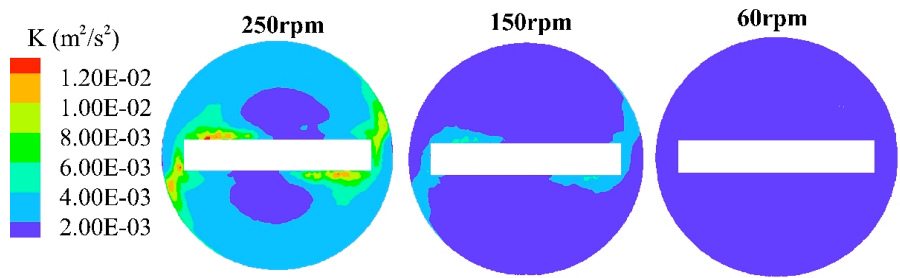

Fig. 5. Distribution of turbulent kinetic energy on a radial plane at the middle of balde (z=15 mm).

The distribution of turbulent kinetic energy on a radial plane at the middle of blade (z=15 *mm*) is shown in Fig. 5. For the stirring speed of 250 *rpm*, the turbulent kinetic energy near the blade tips is about 0.012 $m^2/s^2$, which is almost six times larger than that at the central area. It is because the flow speed in the gap between blade tips and wall boundary is relatively high, which inherently leads to larger Reynolds number, and of course, stronger turbulence intensity near the blade tips. We can also find that the turbulent kinetic energy is nearly constant (0.004 $m^2/s^2$) at the marginal region. This result coincides well with the previous study of Vakili & Esfahany[18]. The effect of stirring speed is also significant on the distribution of turbulent kinetic energy. For the stirring speed of 150 *rpm*, the turbulent kinetic energy is only about one half of that for 250 *rpm*, and it decreases to a nearly constant value of about 0.002 $m^2/s^2$ for 60 *rpm*. This result implies that the turbulent intensity that is highly relevant with the mixing process and the membrane fouling can be obviously modified by the stirring speed.

**3.2 MEMBRANE FOULING EVOLUTION**

The container is initially filled by pure water. We consider the feed solution is pumped into the container from inlet at the beginning of numerical simulations. The feed solution, which consists of a small amount (1% volumetric fraction) of granular materials dispersed in pure water, is transported and mixed by the fluid flow in a well-stirred reactor. This granular (fluid-solid) flow has been carefully investigated to investigate the mixing process of pollutant materials and the evolution of membrane fouling. The mixing process of granular materials in a container under different stirring speeds is shown in Fig. 6. The granular materials entering into the container are transported by the pressure-driven flow and move towards the membrane surface. Under the effect of membrane filtration, the volumetric fraction of granular materials in container continues to accumulate. Under a low stirring speed (60 *rpm*), we can find in Fig. 6 (a) that, the granular materials take a non-uniform distribution on the vertical plane at the initial flow stage (20 *s*). As the circulation loops and vortex could increase the homogeneity of pollutant distribution, the granular materials expand both radially and vertically as time passes, the granular materials in the container distribute more and more symmetrically. Under high stirring speeds (150 and 250 *rpm*), the distribution of granular materials on the vertical plane is almost symmetric even at the initial flow stage (20 *s*). It implies that the stirring speed not only can increase the flow velocity, but also can accelerate the mixing process in the container. Although the granular materials could expand both axially and vertically, it is full of interest to notice that, the region of high concentration of granular materials is always close to the shaft. It mainly comes as a result of flow characteristics, as discussed in section 3.1, the low flow velocity at the near-shaft region allows the granular materials to stay for a long time.

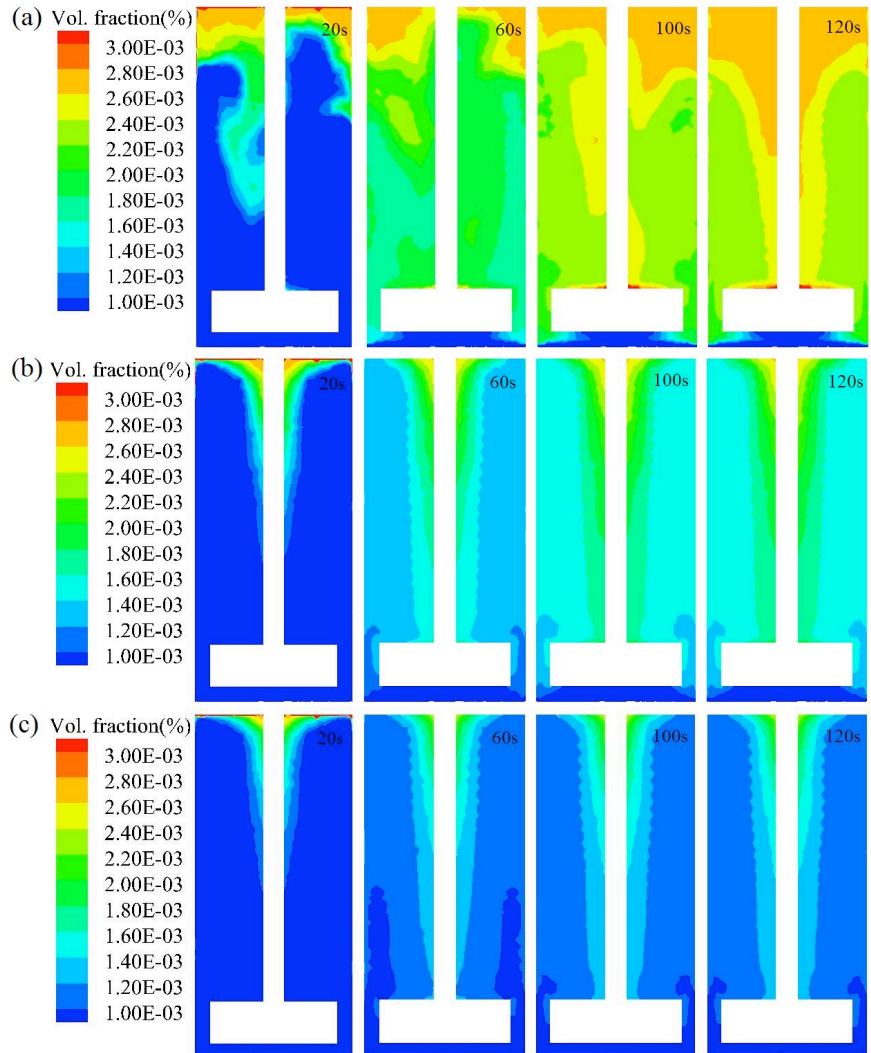

Fig. 6. Mixing evolution of granular materials in a container at different stirring speeds. The stirring speed is (a) 60 rpm, (b) 150 rpm and (c) 250 rpm, respectively.

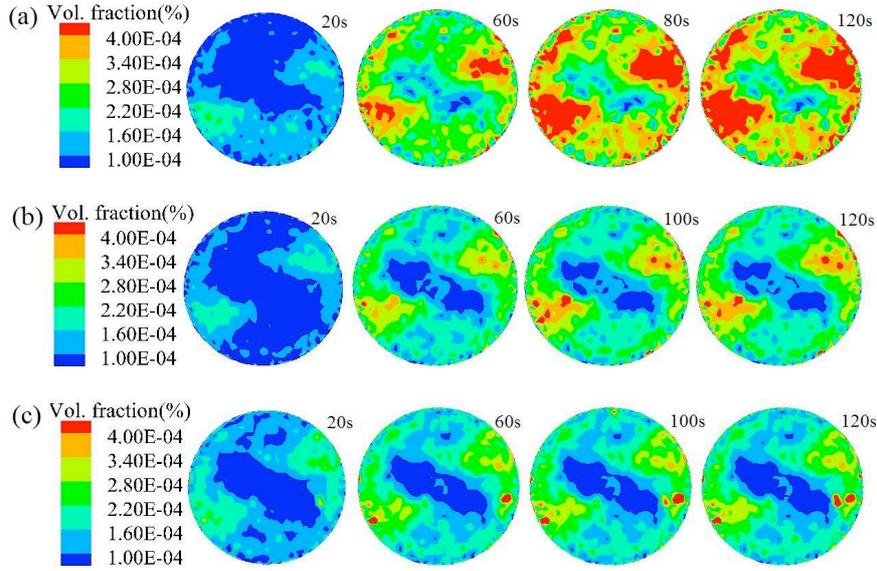

Fig. 7. Evolution of membrane fouling at different stirrring speeds. The stirring speed is (a) 60 rpm, (b) 150 rpm and (c) 250 rpm, respectively.

Fig. 7 shows the evolution of membrane fouling at various stirring speeds. At the initial flow stage (20 *s*), the volumetric fraction of granular materials on the membrane surface is small, but it increases remarkably with time process. Our numerical results have successfully explained in a qualitative way that, the evolution of membrane fouling is accompanied and promoted by the adhesion of granular materials on the membrane surface. The volumetric fraction of granular materials is found to be relatively low at the membrane center, which could imply that the central area of membrane plays a negligible role in the membrane separation process. According to the comparative analysis in Fig. 7 (a-c), we can find that an accelerated stirring speed can significantly decrease the volumetric fraction of granular materials on the membrane surface. This result has theoretically confirmed the experimental fact that the utilization of high stirring speeds is a practical strategy to avoid or to control the membrane fouling effect.

The time evolution of average volumetric fraction of granular materials on the membrane surface is presented in Fig. 8. The granular materials are found to deposit constantly on the membrane surface as time passes, which can be the factor that leads to the formation and development of membrane fouling. At the initial stage (< 20 *s*), the average volumetric fraction of granular materials increases more aggressively for a higher stirred speed. This is because a higher stirring speed could accelerate the transportation of granular materials from inlet to outlet (membrane). However, when the initial stage is passed (> 20 *s*), the increasing of stirring speed can significantly decrease the temporal deposition of granular materials on the membrane surface.

## 4. CONCLUSIONS

This paper focuses on the numerical study of flow characteristic and membrane fouling process in a widely used enzymatic membrane reactor. A three-dimensional Euler-Euler model, supplemented by the *k-ε* mixture turbulence model and a classical drag function, is derived to describe the turbulent granular flow. The validity of our numerical model is proven by comparisons with experimental results.

In the numerical simulations, we have demonstrated that the characteristic of single-phase flow can be obviously modified by the stirring speed. The increase of stirring speed could not only enlarge the circulation loops but also induce significant growth of rotational velocity, shear stress and turbulent kinetic energy in the reactor. By the study of two-phase flow, we find that the mixing process of granular materials is accelerated by the increase of stirring speed. Furthermore, we have explained the membrane fouling mechanism, in a qualitative way, by applying inspection to the time evolution of volumetric function of granular materials on the membrane surface. As these numerical predications coincide well with experimental observations, we could conclude to a certain extent that our numerical model is appropriate for the simulations of membrane fouling effect. However, further studies are still needed for the investigations in reactors of more complex geometry.

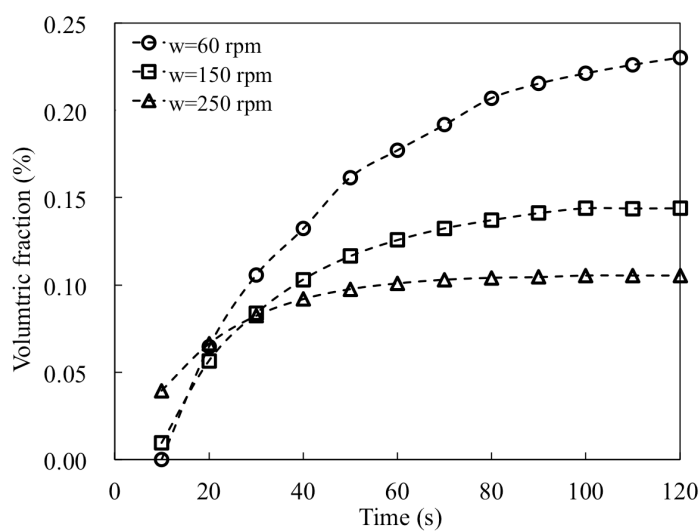

Fig. 8. Time evolution of average volumetric fraction of granular materials on the membrane surface.

**ACKNOWLEDGMENTS**

This work is supported by the grants from the National Natural Science Foundation of China (No. 11402084 & 21506229), the Natural Science Foundation of Hunan Province (No. 2015JJ3051), the self-determined project of State Key Laboratory of Advanced Design and Manufacturing for Vehicle Body (No. 51475002) and the Fundamental Research Funds for the Central Universities (Hunan University).